\documentstyle[amssymb,preprint,aps,epsf,floats]{revtex}

\def\overleftrightarrow{\raise1.5ex\hbox{$\leftrightarrow$}\mkern-16.5mu}

\newcommand{\lia}{$L_{1,A}\ $}

\tighten

\tighten
\preprint{\vbox{
\hbox{DOE/ER/40762-265}
\hbox{MIT-CTP-3312}
}} \bigskip \bigskip

\begin{document}
\title{Constraining the Leading Weak Axial Two-body Current by SNO 
and Super-K }
\author{Jiunn-Wei Chen}
\address{Center for Theoretical Physics,
Massachusetts Institute of Technology, Cambridge, MA 02139\\
and Department of Physics, University of Maryland, College Park,
MD 20742\\
{\tt jwchen@lns.mit.edu}}
\author{Karsten M. Heeger}
\address{Lawrence Berkeley National Laboratory, Berkeley, CA 94720\\
and Department of Physics and Center for Experimental Nuclear Physics
and Astrophysics, \\
University of Washington, Seattle, WA 98195 \\
{\tt KMHeeger@lbl.gov}}
\author{R.G. Hamish Robertson}
\address{Department of Physics and Center for Experimental Nuclear 
Physics and Astrophysics, \\
University of Washington, Seattle, WA 
98195 \\
{\tt rghr@u.washington.edu}}
\maketitle

\begin{abstract}
We analyze the Sudbury Neutrino Observatory (SNO) and 
Super-Kamiokande (SK) data on charged current (CC), neutral current 
(NC) and neutrino electron elastic scattering (ES) reactions to 
constrain the leading weak axial two-body current parameterized by 
$L_{1,A}.$ This two-body current is the dominant uncertainty of
every low energy weak interaction deuteron breakup process,
including SNO's CC and NC reactions. Our method shows that the 
theoretical inputs to SNO's determination of the CC and NC fluxes
can be self-calibrated, be calibrated by SK, or be calibrated by 
reactor data. The only assumption made is that the total flux of active 
neutrinos has the standard $^{8}B$ spectral shape (but distortions in 
the electron neutrino spectrum are allowed). We show that SNO's
conclusion about the inconsistency of the no-flavor-conversion 
hypothesis does not contain significant theoretical uncertainty, and 
we determine the magnitude of the active solar neutrino flux.
\end{abstract}

\preprint{\vbox{
\hbox{ } }} \bigskip \bigskip \vfill\eject

\section{Introduction}

Recent conclusive results from the Sudbury Neutrino Observatory~(SNO) have
established the existence of non-electron active neutrino components in the $%
^{8}B$ solar neutrino flux \cite{SNO1} and hence have given a strong
evidence for neutrino oscillation. These results are based on the three
reactions measured by SNO to detect the $^{8}B$ solar flux 
\begin{equation}
\begin{array}{llll}
\nu _{e}+d & \rightarrow & p+p+e^{-} & \qquad \text{(CC)}, \\ 
\nu _{x}+d & \rightarrow & p+n+\nu _{x} & \qquad \text{(NC)}, \\ 
\nu _{x}+e^{-}\!\!\!\! & \rightarrow & \nu _{x}+\text{{}}e^{-} & \qquad 
\text{(ES).}
\end{array}
\end{equation}
\ The charged current reaction (CC) is sensitive exclusively to
electron-type neutrinos, while the neutral current reaction (NC) is equally
sensitive to all active neutrino flavors ($x=e,\mu ,\tau $). The elastic
scattering reaction (ES) is sensitive to all active flavors as well, but
with reduced sensitivity to $\nu _{\mu }$ and $\nu _{\tau }$. Detection of
these three reactions allows SNO to determine the electron and non-electron
active neutrino components of the solar flux, and it is then obvious that
the cross sections for these three reactions are important inputs for SNO.
The cross sections for all three reactions are determined from theory, but
the CC and NC cross sections involve nuclear-physics complexities not
present in the ES interaction description. Thus the CC and NC cross sections
have become the main source of theoretical uncertainties for SNO.

The complexities in the CC and NC processes are due to two-body currents
which are interactions involving two nucleons and external leptonic
currents. In the potential model approach, the two-body currents are
associated with the meson exchange currents and can be calculated in terms
of unknown weak couplings. In effective field theory (EFT), the two-body
currents are parameterized. In both cases, experimental data from some other
processes are required in order to calibrate the unknowns in the problem. In
EFT, this calibration procedure can be described in an economic and
systematic way. The reason is that, up to next-to-next-to-leading order
(NNLO) in EFT, all low-energy weak interaction deuteron breakup processes
depend on a common isovector axial two-body current, parameterized by $%
L_{1,A}$ \cite{BCK} (see more explanations in the next section). This
implies that a measurement of any one of the breakup processes could be used
to fix $L_{1,A}$. A summary of the previous efforts in the determination of $%
L_{1,A}$ can be found in Ref. \cite{BCV}.

In this paper, after briefly reviewing the EFT approach, we will present the
constraint on $L_{1,A}$ using a combined analysis of the CC, NC and ES data
from SNO\ and Super-Kamiokande (SK). We then compare this new result with
other determinations of $L_{1,A}$ and comment on the interpretation of SNO's
measurements with the assumption about the size of $L_{1,A}$ eliminated.

\section{Effective Field Theory}

For the deuteron breakup processes used to detect solar neutrinos, where the
neutrino energies $E_{\nu }<15$~MeV, the typical momentum scales in the
problem are much smaller than the pion mass $m_{\pi }(\simeq 140$ MeV$).$ In
these systems pions do not need to be treated as dynamical particles since
they only propagate over distances $\sim 1/m_{\pi }$, much shorter than the
scale set by the typical momentum of the problem. Thus the pionless nuclear
effective field theory, ${\rm EFT}({\pi \hskip-0.6em/})$ \cite
{KSW96,K97,vK97,Cohen97,BHvK1,CRS}, is applicable.

In ${\rm EFT}({\pi \hskip-0.6em/})$, the dynamical degrees of freedom are
nucleons and non-hadronic external currents. Massive hadronic excitations
such as pions and the delta resonance are not dynamical. Their contributions
are encoded in the contact interactions between nucleons. Nucleon-nucleon
interactions are calculated perturbatively with the small expansion
parameter 
\begin{equation}
Q\equiv \frac{\left( 1/a,\gamma ,p\right) }{\Lambda }
\end{equation}
which is the ratio of the light to heavy scales. The light scales include
the inverse S-wave nucleon-nucleon scattering length $1/a(\lesssim 12$ MeV$)$
in the $^{1}S_{0}$ channel, the deuteron binding momentum $\gamma (=45.7$
MeV) in the $^{3}S_{1}$ channel, and the typical nucleon momentum $p$ in the
center-of-mass frame. The heavy scale $\Lambda $ is set by the pion mass $%
m_{\pi }$. This formalism has been applied successfully to many processes
involving the deuteron~\cite{CRS,npdgam2}, including Compton scattering \cite
{dEFT,GR}, $np\rightarrow d\gamma $ for big-bang nucleosynthesis \cite
{npdgam1,Rupak}, $\nu d$ reactions for SNO physics \cite{BCK}, the solar $pp 
$ fusion process \cite{KR,pp}, and parity violating observables \cite{PV} .
In addition, this formalism has been applied successfully to three-nucleon
systems \cite{BHvK2}. It has revealed highly non-trivial renormalizations
associated with three body forces in the $s_{1/2}$ channel (e.g., $^{3}$He
and the triton). For other channels, precision calculations were carried out
to higher orders \cite{BHvK1}.

For low energy deuteron breakup processes, it is well known that the
dominant contributions to the hadronic matrix elements are the $%
^{3}S_{1}\rightarrow $ $^{1}S_{0}$ transitions through the isovector axial
couplings. The $^{3}S_{1}$ state (such as a deuteron) has spin $S=1$ and
isospin $I=0$, while the $^{1}S_{0}$ state has $S=0$ and $I=1$. Amongst the
spin-isospin operators {\bf 1}, $\tau ^{a}$, $\sigma ^{i}$ and $\tau
^{a}\sigma ^{i}$, only the isovector axial coupling $\tau ^{a}\sigma ^{i}$
can connect $^{3}S_{1}$ to $^{1}S_{0}$ states. The $^{3}S_{1}\rightarrow
\,^{3}S_{1}$ transitions are suppressed at low energies because i) the
isovector operators do not contribute (the transition is isoscalar) and ii)
the matrix elements of the one-body isoscalar operators vanish in the zero
recoil limit ($d$ and $np$ states are orthogonal in this limit). This leads
to large suppression of the isoscalar two-body contributions through the
interference terms. Also, at low energies, the non-derivative operators are
more important than the derivative operators. Thus the leading two-body
current contributions for low energy weak interaction deuteron breakup
processes only depend on a non-derivative, isovector axial two-body current, 
$L_{1,A}$.

In Ref.~\cite{BCK}, ${\rm EFT}({\pi \hskip-0.6em/})$ is applied to compute
the cross-sections for four channels (CC, NC, $\overline{\nu }%
_{e}+d\rightarrow e^{+}+n+n$ and $\overline{\nu }_{x}+d\rightarrow \overline{%
\nu }_{x}+n+p$) to NNLO, up to 20 MeV (anti)neutrino energies. As already
mentioned, these processes have been shown to depend on only one parameter, $%
L_{1,A}$. This dependence is subject to an intrinsic uncertainty in our EFT
calculation at NNLO of less than 3\%. Through varying $L_{1,A}$, the
potential model results of Refs.~\cite{YHH} and \cite{NSGK} are reproduced
to high accuracy for all four channels. This confirms that the $\sim 5\%$
difference between Refs.~\cite{YHH} and \cite{NSGK} is due largely to
different assumptions made about short distance physics.

The same two-body current $L_{1,A}$ also contributes to the proton-proton
fusion process $p+p\rightarrow d+e^{+}+\nu _{e}$. This is the primary
reaction in the $pp$ chain of nuclear reactions that power the sun,
reactions which in turn generate the neutrino flux to be observed by SNO.
The calculations in ${\rm EFT}({\pi \hskip-0.6em/})$ were carried out
initially to second order \cite{KR}, and then to fifth order \cite{pp}. Thus
a calibration to SNO's CC and NC reactions can also be used to calibrate the
proton-proton fusion process.

\section{Fixing $L_{1,A}$ From a Combined NC, CC and ES Analysis}

In this section we present the constraint on $L_{1,A}\ $obtained from a
combined analysis of the solar neutrino fluxes measured by CC, NC, and ES
reactions. In SNO's analysis, a specific $L_{1,A}\ $was chosen in CC and NC
reactions. The extracted solar neutrino fluxes from CC and NC were then
compared to each other and to ES to extract a consistent set of neutrino
flavor-conversion probabilities and to map allowed regions in a 2-mass
mixing description. Here we take $L_{1,A}\ $as a free parameter and use the
available experimental data from SNO and SK to fix not only the
flavor-conversion probabilities but also $L_{1,A}$. The only assumption we
will make is that the {\em total} flux for the active solar neutrinos has
the standard $^{8}$B shape.

In the two-flavor oscillation analysis there are three parameters extracted, 
$\Delta m_{12}^{2}$, $\theta _{12}$, and $\Phi _{\nu _{x}}$, which are the
differences between the squares of the neutrino masses, the mixing angle,
and the total active neutrino flux. There are three separate experimental
inputs, the ES, CC, and NC rates. It might at first be thought impossible to
extract a fourth parameter, $L_{1,A}$, without additional inputs or
assumptions, such as fixing the shape of the electron-neutrino spectrum. The
shape is experimentally determined, but not yet with high accuracy. Our
strategy is to note that, in active-only oscillations, there is no shape
distortion in the total flux, and that the integrated spectral response in
the CC reaction over a certain range of final electron energies is the same
as that in the ES reaction over a different range of energies, independent
of distortions of the neutrino spectrum \cite{VFL}.

The CC and NC are measured at SNO and the ES is both measured at SNO and SK.
The measured event rates are the integrals of the effective cross sections
weighted by the solar neutrino fluxes that reached the target.

\begin{eqnarray}
R_{NC} &=&\int dE\widetilde{\sigma }_{NC}f_{\nu _{x}}\ ,  \label{m1a} \\
R_{CC} &=&\int dE\widetilde{\sigma }_{CC}f_{\nu _{e}}\ ,  \label{m1b} \\
R_{ES} &=&\int dE\widetilde{\sigma }_{e}f_{\nu _{e}}+\int dE\widetilde{%
\sigma }_{\mu ,\tau }\left[ f_{\nu _{x}}-f_{\nu _{e}}\right] \ ,  \label{m1c}
\end{eqnarray}
where $R_{i}$ is the event rate, $f_{\nu _{i}}$ is the $\nu _{i}$ flux and $%
E $ is the neutrino energy. $\widetilde{\sigma }_{i}$ is the effective cross
section, defined in Appendix A, with $i=e,\mu ,\tau $ for $\nu _{e,\mu ,\tau
}+e$ ES interaction. These effective cross sections are the true cross
sections convoluted with the detector resolution functions which describe
how the energy is transferred to electrons and detected by their Cherenkov
radiations. The effective cross sections depend on the electron detection
threshold $T_{i}^{th}$. For CC and NC reactions, they also depend on $%
L_{1,A} $.

The total flux for the active solar neutrinos is assumed to have the
standard $^{8}$B shape, 
\begin{equation}
f_{\nu _{x}}\left( E\right) =\Phi _{\nu _{x}}\phi _{^{8}B}\left( E\right) \ ,
\label{a1}
\end{equation}
where $\phi _{^{8}B}$ is the normalized $^{8}B$ shape function ($%
\int_{0}^{E_{\max }}dE\phi _{8B}\left( E\right) =1$) \cite{8B} and $\Phi
_{\nu _{x}}$ is the magnitude of the $\nu _{x}$ flux. This assumption is
valid if there are no oscillations to sterile neutrinos, or, even if such
mixing {\em is} present, the survival probability to active neutrinos is
energy independent. Similarly, the $\nu _{e}$ flux is 
\begin{equation}
f_{\nu _{e}}\left( E\right) =\Phi _{\nu _{x}}\phi _{^{8}B}\left( E\right)
P_{\nu _{e}/\nu _{x}}\left( E\right) \ ,  \label{a2}
\end{equation}
where $P_{\nu _{e}/\nu _{x}}\left( E\right) $ is the probability
distribution of finding a $\nu _{e}$ out of a $\nu _{x}$. Obviously, $P_{\nu
_{e}/\nu _{x}}\left( E\right) $ is bounded between $0$ and $1$.

We follow Villante et al. in Ref.\cite{VFL} to define the averaged effective
cross sections $\overline{\sigma }_{i}$ and the normalized response
functions $\rho _{i}\left( E\right) $ for the $^{8}B$ spectrum 
\begin{eqnarray}
\overline{\sigma }_{i} &\equiv &\int dE\widetilde{\sigma }_{i}\phi
_{^{8}B}\left( E\right) \ ,  \nonumber \\
\rho _{i}\left( E\right) &=&\frac{\widetilde{\sigma }_{i}\phi _{^{8}B}\left(
E\right) }{%
\displaystyle\int %
dE\widetilde{\sigma }_{i}\phi _{^{8}B}\left( E\right) }\ .  \label{a4}
\end{eqnarray}
Then we rewrite eqs. (\ref{m1a}-\ref{m1c}) using eqs.(\ref{a1}-\ref{a4})

\begin{eqnarray}
R_{NC} &=&\Phi _{\nu _{x}}\overline{\sigma }_{NC}\ ,  \label{m2a} \\
R_{CC} &=&\Phi _{\nu _{x}}\overline{\sigma }_{CC}\int dE\rho _{CC}P_{\nu
_{e}/\nu _{x}}\ ,  \label{m2b} \\
R_{ES} &=&\Phi _{\nu _{x}}\overline{\sigma }_{\mu ,\tau }+\Phi _{\nu
_{x}}\int dE\left[ \overline{\sigma }_{e}\rho _{e}-\overline{\sigma }_{\mu
,\tau }\rho _{\mu ,\tau }\right] P_{\nu _{e}/\nu _{x}}\ .  \label{m2c}
\end{eqnarray}
An important observation made in Ref. \cite{VFL} is that the terms
with the
normalized response function $\rho \left( E\right) $ dependence can be
related by choosing suitable detection thresholds. This allows us to reduce
the numbers of unknowns such that eqs.(\ref{m2a}-\ref{m2c}) become solvable.
These approximate relations and their corrections are systematically
explored in Appendix A. As shown in Fig.1(b), $\rho _{CC}$ is very
insensitive to $L_{1,A}$. Thus we can set 
\begin{equation}
\frac{\partial }{\partial L_{1,A}}\rho _{CC}\left( E\right) |_{T_{CC}^{th}=5%
\text{MeV}}=0\ .  \label{s}
\end{equation}
Also, as shown in Fig. 1(c), to a very good approximation, 
\begin{equation}
\rho _{e}\left( E\right) |_{T_{e}^{th}=6.8\text{MeV}}=\rho _{\mu ,\tau
}\left( E\right) |_{T_{\mu ,\tau }^{th}=6.8\text{MeV}}\ .  \label{ss}
\end{equation}
The corrections of the above relations change $L_{1,A}$ by up to a
negligible amount of 0.25 fm$^{3}$. A more significant correction comes from 
\begin{equation}
\rho \left( E\right) \equiv \rho _{CC}\left( E\right) |_{T_{CC}^{th}=5\text{%
MeV}}\cong \rho _{e}\left( E\right) |_{T_{e}^{th}=6.8\text{MeV}}\ .
\label{sss}
\end{equation}
We will use a parameter $\epsilon $ in eq.(\ref{m3c}) to parametrize the
correction.

Because of eq.(\ref{s}), the $L_{1,A}$ dependence only shows up in $%
\overline{\sigma }_{NC}$ and $\overline{\sigma }_{CC}$. The scaling can be
written as 
\begin{eqnarray}
\overline{\sigma }_{NC} &=&\overline{\sigma }_{NC}^{0}g_{NC}(L_{1,A})\ , 
\nonumber \\
\overline{\sigma }_{CC} &=&\overline{\sigma }_{CC}^{0}g_{CC}(L_{1,A})\ ,
\label{n2}
\end{eqnarray}
where the $\overline{\sigma }_{NC}^{0}$ and $\overline{\sigma }_{CC}^{0}$
are the values used by SNO \cite{SNO1} which are based on the calculation of
Ref. \cite{SAT+} with the electromagnetic radiative corrections of Ref. \cite
{EM} (see Ref. \cite{BS} for an earlier attempt)
and a 5 MeV electron detection threshold. These cross sections are
corresponding to the NNLO EFT results with $L_{1,A}(\overline{\mu }=m_{\pi
})=4.0$ fm$^{3}$, where the renormalization scale $\overline{\mu }$ is set
to the pion mass. In the following expressions, we will suppress the $%
\overline{\mu }$ dependence of $L_{1,A}$ for simplicity. The scaling
functions can be\ parametrized as 
\begin{eqnarray}
g_{NC}(L_{1,A}) &=&1+\alpha _{NC}\left( \frac{L_{1,A}}{\text{fm}^{3}}%
-4.0\right) \ ,  \nonumber \\
g_{CC}(L_{1,A}) &=&1+\alpha _{CC}\left( \frac{L_{1,A}}{\text{fm}^{3}}%
-4.0\right) \ ,  \label{n3}
\end{eqnarray}
where $\alpha _{NC}=0.013$ and $\alpha _{CC}=0.010$. It is interesting to
note that if $\overline{\sigma }$ were averaged true cross sections instead
of the averaged {\em effective} cross sections, then $\alpha _{NC}$ and $%
\alpha _{CC}$ would be almost identical \cite{BCK}. The larger difference
here is due to the difference in the detection methods. As shown in eqs.(\ref
{2}) and (\ref{3}), the CC event detection requires the final state electron
energy to be above a certain detection threshold, thus leptons transferring
too much energy to the hadrons will not be detected. For NC detection,
however, there is no such discrimination---all the neutrons generated in NC
have the same probability to be detected in the thermalization and capture
process. Thus in general, the scaling factor of NC is associated with that
of more energetic scattering (with larger energy transfer to the hadrons)
than that of CC. We find the qualitative difference between $\alpha _{NC}$
and $\alpha _{CC}$ is consistent with the neutrino energy dependence of the
scaling factors calculated in Ref. \cite{BCK}.

Substituting eqs.(\ref{s}-\ref{n2}) into the eqs. (\ref{m2a}-\ref{m2c}), we
have 
\begin{eqnarray}
R_{NC} &=&\Phi _{\nu _{x}}\overline{\sigma }_{NC}^{0}g_{NC}(L_{1,A})\ ,
\label{m3a} \\
R_{CC} &=&\Phi _{\nu _{x}}\overline{\sigma }_{CC}^{0}g_{CC}(L_{1,A})\int
dE\rho P_{\nu _{e}/\nu _{x}}\ ,  \label{m3b} \\
R_{ES} &=&\Phi _{\nu _{x}}\overline{\sigma }_{e}\left[ \frac{\overline{%
\sigma }_{\mu ,\tau }}{\overline{\sigma }_{e}}+\left( 1-\frac{\overline{%
\sigma }_{\mu ,\tau }}{\overline{\sigma }_{e}}\right) \left( 1+\epsilon
\right) \int dE\rho P_{\nu _{e}/\nu _{x}}\right] \ .  \label{m3c}
\end{eqnarray}
Note that $T_{CC}^{th}=T_{NC}^{th}=5$ MeV and $T_{\mu ,\tau
}^{th}=T_{e}^{th}=6.8$ MeV to be consistent with eq.(\ref{sss} ). $\epsilon $
parametrizes the correction to the approximate identity of eq.(\ref{sss}).
As shown in Appendix A , there is a model independent bound 
\begin{equation}
\left| \epsilon \right| <4\%\ .
\end{equation}
This correction 
introduces changes $L_{1,A}$ by up to $\pm 2.0\ $fm$^{3}$ but
changes
the other quantities by negligible amounts.

Now it is clear that eqs.(\ref{m3a}-\ref{m3c}) are solvable. The three
equations determine three quantities: $\Phi _{\nu _{x}}$, $L_{1,A}$ and $%
\int dE\rho P_{\nu _{e}/\nu _{x}}$, which are the magnitude of the total
active neutrino flux, the axial two-body current, and the measured $\nu
_{e}/\nu _{x}$ ratio, respectively. If there is no neutrino oscillation,
then $\int dE\rho P_{\nu _{e}/\nu _{x}}=1$.

For the experimental inputs, SNO has measured the ES rates, but the SK
determination is more precise while in agreement with SNO. Therefore we use
the SK measurements not only to provide a value for the integral above 6.8
MeV as shown above, but also to fix (and remove) the ES contribution to the
total SNO rates above 5 MeV. With 1496 days of data, SK reports \cite{SK}
the equivalent electron neutrino fluxes ($\equiv R_{ES}/\overline{\sigma }%
_{e}$) for analysis thresholds of 6.5 and 5.0 MeV, 
\begin{eqnarray}
\Phi _{{\rm SK}}(6.5) &=&\left( 2.362_{-0.068}^{+0.074}\right) \times 10^{6}%
{\rm \ cm}^{-2}{\rm s}^{-1}\ , \\
\Phi _{{\rm SK}}(5.0) &=&\left( 2.348_{-0.066}^{+0.073}\right) \times 10^{6}%
{\rm \ cm}^{-2}{\rm s}^{-1}\ ,
\end{eqnarray}
respectively, where the statistical and systematic errors have been added in
quadrature. Here and in the subsequent analysis in this paper, where
asymmetric errors occur, we simply use the larger. We take the uncertainties
in the two SK fluxes to be fully correlated. In view of the lack of
significant threshold-energy dependence in the SK flux, we assume 
\begin{equation}
\Phi _{{\rm SK}}(6.8)=\Phi _{{\rm SK}}(6.5)\ .
\end{equation}

SNO provides a model-independent value for the NC rate, obtained by
separating the CC and NC parts of the signal by their different radial and
sun-angle dependences in the detector. This result is independent of the
energy spectrum of the CC events. Converting the flux reported into an
equivalent number of events gives $727\pm 190$ detected in the 306.4-day
running period.

\begin{table}[tbp]
\caption{Numbers of events reported by SNO for 306.4 live days for a kinetic
energy threshold of 5 MeV.  The uncertainties (except for the backgrounds) are statistical.}
\label{SNO02}
\begin{center}
\begin{tabular}{lrc}
Reaction & Events & Uncertainty \\ \hline
Candidate Events & 2928 & 54.1 \\ 
Backgrounds & 123 & +21.6 -17.0 \\ 
Total Neutrino Events & 2805 & +58.3 -56.7 \\ 
ES (from SK) & 258.3 & 8.0 \\ 
Net NC + CC & 2546.7 & 59 \\ 
NC (CC shape unconstrained) & 727 & 190 \\ 
NC (CC shape constrained) & 576.5 & 49.5
\end{tabular}
\end{center}
\end{table}
%
%

In Table \ref{SNO02} the event rates needed for the model-independent
analysis are summarized. The ``true'' number of ES events in the SNO data
set is derived from $\Phi _{{\rm SK}}(5.0)$ and the SNO effective elastic
scattering cross section with a 5.0 MeV threshold ${\overline{\sigma }_{e}|}%
_{5.0}$, 
\begin{equation}
{\overline{\sigma }_{e}|}_{5.0}=1.10\times 10^{-4}\text{\ cm}^{2}\text{sT}%
^{-1}\ ,
\end{equation}
derived from Ref. \cite{SNO1}. (The number is in excellent agreement with
the $263.6\pm 26.4({\rm stat.})$ obtained during the SNO signal extraction.)
One could derive a value for the CC rate directly from the fifth and sixth
lines of this table, but the two would be highly correlated. It is
preferable to make use of expressions for NC + CC and NC because the summed
rate is essentially free of correlation with the NC rate. So we use 
\begin{eqnarray}
R_{NC}+R_{CC} &=&R_{tot}-\Phi _{{\rm SK}}(5.0){\overline{\sigma }_{e}|}_{5.0}%
{=}\left( 2546\pm 59\right) \text{T}^{-1}\text{\ ,}  \nonumber \\
R_{NC} &=&\left( 727\pm 190\right) \text{T}^{-1}\ ,  \label{Yo}
\end{eqnarray}
where T$=$306.4 days.

The averaged effective cross sections of SNO can be extracted from Ref. \cite
{SNO1} 
\begin{eqnarray}
\overline{\sigma }_{NC}^{0} &=&1.13\times 10^{-4}\text{\ cm}^{2}\text{sT}%
^{-1}\ ,  \nonumber \\
\overline{\sigma }_{CC}^{0} &=&1.12\times 10^{-3}\text{\ cm}^{2}\text{sT}%
^{-1}\ .
\end{eqnarray}
The effective cross sections are subject to uncertainty from a variety of
sources, tabulated by SNO \cite{SNO1}. These include principally the energy
scale, vertex-reconstruction accuracy, and (for the NC reaction) energy
resolution and neutron capture efficiency. The sources of uncertainty
produce in some cases correlated variations in the effective cross sections,
which are explicitly accounted for in the analysis.

In general we should have added $3\%$ systematics for the NNLO EFT
calculations of $\overline{\sigma }_{NC}^{0}$ and $\overline{\sigma }%
_{CC}^{0}$, because for an EFT with a small expansion parameter $Q\sim 1/3$,
3\% error is reasonable for a third order (NNLO) calculation. In the
analysis of \cite{BCK}, however, a faster convergence is seen in four
channels of (anti)neutrino-deuteron scattering, such that 1-2\% higher
corrections also seems reasonable. Furthermore, it is conceivable that the
higher order corrections can be absorbed in $L_{1,A}$ in low energy
processes. One indication that this might happen is in the comparison with
the potential model calculations. The potential model results have quite
different systematics to those of EFT. The fact that NNLO EFT can fit four
channels of (anti)neutrino-deuteron reaction results of \cite{NSGK} to
within $1\%$ \cite{BCK} suggests that higher order effects can be absorbed
in $L_{1,A}$. Further investigation is still required to see whether the
higher order effects shift $L_{1,A}$ approximately the same amount. For
matrix elements with similar kinematics, this is likely to be true. In our
case, we have CC and NC in approximately the same energy region. Thus we
expect the higher order effects just shift $L_{1,A}$ by a certain amount ($%
\sim +2$ to $+3$ fm$^{3}$, with the sign fixed by the fifth order proton-proton
fusion calculation \cite{pp,BCV} as will be explained in more detail later)
without introducing additional error to the $\Phi _{\nu _{x}}$ and $\int
dE\rho P_{\nu _{e}/\nu _{x}}$ determinations.

For ES reactions with a $6.8$-MeV threshold, the ratio of the neutral
current and electron neutrino scattering cross sections is, 
\begin{equation}
\frac{\overline{\sigma }_{\mu ,\tau }}{\overline{\sigma }_{e}}=0.153\ ,
\end{equation}
with radiative corrections included.

Now we have all the inputs required to solve eqs. (\ref{m3a}-\ref{m3c}). The
full set of equations is nonlinear in $L_{1,A}$, but a linearized solution
may be obtained by making a first-order expansion for $g_{CC}/g_{NC}=1+(%
\alpha _{CC}-\alpha _{NC})(L_{1,A}/$fm$^{3}-4.0)+\ldots $ The term quadratic
in \ $(L_{1,A}/$fm$^{3}-4.0)$ is $\sim 10^{-4}$ and can be neglected. Using
this approximation, the solutions of eqs.(\ref{m3a}-\ref{m3b}) are: 
\begin{eqnarray}
(\frac{L_{1,A}}{\text{fm}^{3}}-4.0) =\left[ \alpha _{CC}\Phi _{{\rm SK}}(6.8)%
\overline{\sigma }_{CC}^{0}+(\alpha _{NC}-\alpha _{CC})R_{NC}\frac{\overline{%
\sigma }_{\mu \tau }}{\overline{\sigma }_{e}}\frac{\overline{\sigma }%
_{CC}^{0}}{\overline{\sigma }_{NC}^{0}}\right] ^{-1}  \nonumber \\
\times \left\{ \left[ R_{tot}-\Phi _{{\rm SK}}(5.0){\overline{\sigma }_{e}|}%
_{5.0}-R_{NC}\right] \left( 1-\frac{\overline{\sigma }_{\mu \tau }}{%
\overline{\sigma }_{e}}\right) (1+\epsilon )-\Phi _{{\rm SK}}(6.8){\overline{%
\sigma }_{CC}^{0}}+R_{NC}\frac{\overline{\sigma }_{\mu \tau }}{\overline{%
\sigma }_{e}}\frac{\overline{\sigma }_{CC}^{0}}{\overline{\sigma }_{NC}^{0}}%
\right\} ,
\end{eqnarray}

\begin{equation}
\int dE\rho P_{\nu _{e}/\nu _{x}}=\left( \Phi _{{\rm SK}}(6.8)\frac{g_{NC}%
\overline{\sigma }_{NC}^{0}}{R_{NC}}-\frac{\overline{\sigma }_{\mu ,\tau }}{%
\overline{\sigma }_{e}}\right) \left( 1-\frac{\overline{\sigma }_{\mu ,\tau }%
}{\overline{\sigma }_{e}}\right) ^{-1}\left( 1+\epsilon \right) ^{-1},
\end{equation}

\begin{equation}
\Phi _{\nu _{x}}=\frac{R_{NC}}{g_{NC}\overline{\sigma }_{NC}^{0}}\ .
\end{equation}

Inserting the experimental and theoretical quantities, 
\begin{eqnarray}
L_{1,A} &=&4.0\pm 4.7 (\rm stat.) \pm 4.5 (\rm syst.) \ \text{fm}^{3}\ ,  \nonumber \\
\Phi _{\nu _{x}} &=&\left( 6.4\pm 1.4 \pm 0.6 \right) \times 10^{6}\ \text{cm}^{-2}%
\text{s}^{-1}\ ,  \label{results} \\
\int dE\rho P_{\nu _{e}/\nu _{x}} &=&0.25_{-0.07}^{+0.12} \pm 0.03 \ .  \nonumber
\end{eqnarray}
The statistical errors in \lia are dominated by $R_{NC}$, and the systematic errors by $\Phi_{\rm SK}$ and by vertex reconstruction accuracy in SNO.

%
%
\begin{table}[tbp]
\caption{{Determinations of the NNLO $L_{1,A}$ (at renormalization scale $m_{%
\protect\pi }$) from different processes. The higher order theoretical
systematics are expected to be absorbed by shifting $L_{1,A}$ by $\sim +2$ to $%
+3 $ fm$^{3}$ thus is not included in this table. Note that the CC, NC \& ES
combined analysis assumes the standard $^{8}B$ shape for the active neutrino
flux. The tritium $\protect\beta$ decay analysis assumes the three-body
current is negligible. The helioseismology analysis does not include the
uncertainties from the solar model. The last two entries are theoretical
determinations. EFT dimensional analysis gives $\left|L_{1,A}\right| $ $\sim
6$ (fm$^{3}$) which is denoted as $[-6,6]$ as its expected range. }}
\label{L1a}
\begin{center}
\begin{tabular}{lcr}
Processes & $L_{1,A}$ (fm$^{3}$) & References \\ \hline
CC, NC \& ES & $4.0\pm 6.3$ & [this work] \\ 
Reactor $\overline{\nu }$-$d$ & $3.6\pm 4.6$ & \cite{BCV} \\ 
Tritium $\beta $ decay & $4.2\pm 0.1$ & \cite{Tritium}(see also 
\cite{pp,BCV,park}%
) \\ 
Helioseismology & $4.8\pm 5.9$ & \cite{helio} \\ 
Dimensional analysis & $\sim \lbrack -6,6]$ & \cite{BCK} \\ 
Potential model & $4.0$ & \cite{SAT+}
\end{tabular}
\end{center}
\end{table}
%
%

A few comments can be made about the result we obtain in eq.(\ref{results}).
First, the only assumption we have made is that the active neutrino flux has
the standard $^{8}B$ shape. The rest of the treatment is model independent
in the sense that we have not assumed the size of $L_{1,A}$ or assumed any
neutrino oscillation scenarios. Second, our result on $\int dE\rho P_{\nu
_{e}/\nu _{x}}$ can be used to constrain neutrino oscillation parameters.
Third, the size of the active neutrino flux is consistent with the $\nu _{e}$
flux of the standard solar model $\Phi _{SSM}=\left(
5.05_{-0.81}^{+1.01}\right) \times 10^{6}\ $cm$^{-2}$s$^{-1}$. This sets a
constraint on the oscillations between the active and sterile neutrinos.
Fourth, the range of\ $L_{1,A}$ we have obtained is consistent with the
estimated value $\left| L_{1,A}\right| $ $\sim 6$ fm$^{3}$ (at $\overline{%
\mu }=m_{\pi }$) from dimensional analysis \cite{BCK}. It is also consistent
with the constraints from reactor-antineutrino deuteron breakup processes 
\cite{BCV} , tritium beta decay \cite{Tritium}, helioseismology \cite{helio}%
, and the latest improved potential model results \cite{SAT+} (corresponding
to $L_{1,A}$ $=4.0$ fm$^{3}$). The comparison of their corresponding NNLO $%
L_{1,A}$'s is listed in Table \ref{L1a}. Here we have assumed that most of
the higher order effects in EFT can be absorbed by $L_{1,A}$, and the higher
order theoretical systematics are therefore not included in the assigned
error bars. We expect a $+2$ to $+3$ fm$^{3}$ contribution to the effective
value of $L_{1,A}$ from higher orders. The sign is fixed by an explicit
fifth-order calculation of the proton-proton fusion at threshold \cite{pp}
which shows that $L_{1,A}$ shifts by +2 to +3 fm$^{3}$ from the third order
(NNLO) to the fifth order. Even though the tritium beta decay analysis
assumes that the three-body current is negligible and the helioseismology
analysis does not include the uncertainties from the solar model, it is
still very encouraging that all the constraints agree with each other very
well, given how different the physical systems are.

It is likely in the future the error bar of $R_{NC}$ could be reduced by a
factor of 2. In that case, the error on $L_{1,A}\ $ would be reduced to 5
fm$^{3}$.

 It is also
interesting to reinvestigate the null hypothesis (specifically that all
observed fluxes can be described consistently within the Standard Model 
of Particles and Fields) when $L_{1,A}\ $ is allowed to
float. $\int dE\rho P_{\nu _{e}/\nu
_{x}}=1$ in the Standard Model, and thus the set of
three equations (\ref{m3a}-\ref{m3b}) contain only two parameters. One finds
that the set is inconsistent at 4.3 $\sigma $.   Alternatively, if 
one uses the experimental determination of \lia from reactor data 
(Table \ref{L1a}), the null hypothesis fails at 5.1 $\sigma$ (SNO
only) or 5.3 $\sigma$ (SNO and SK).  Thus, even if SNO were to
place no reliance at all on the theoretical calculations of short-distance
physics \cite{NSGK,SAT+}, it would still be true that the
no-flavor-conversion hypothesis is ruled out with high confidence.

One might suspect that if the value of $L_{1,A}$ is taken from some other
constraints, perhaps the $^{8}B$ shape assumption for the active neutrino
flux can be removed. This question can be easily answered by inspecting the
new set of equations 
\begin{eqnarray}
R_{NC} &=&\Phi _{B}\overline{\sigma }_{NC}^{0}g_{NC}\int dE\rho _{NC}P_{\nu
_{e}\rightarrow \nu _{x}}\ ,  \nonumber \\
R_{CC} &=&\Phi _{B}\overline{\sigma }_{CC}^{0}g_{CC}\int dE\rho P_{\nu
_{e}\rightarrow \nu _{e}}\ , \\
R_{ES} &=&\Phi _{B}\overline{\sigma }_{e}\left[ \frac{\overline{\sigma }%
_{\mu ,\tau }}{\overline{\sigma }_{e}}\int dE\rho _{\mu ,\tau }P_{\nu
_{e}\rightarrow \nu _{x}}+\left( 1-\frac{\overline{\sigma }_{\mu ,\tau }}{%
\overline{\sigma }_{e}}\right) \left( 1+\epsilon \right) \int dE\rho P_{\nu
_{e}\rightarrow \nu _{e}}\right] \ ,  \nonumber
\end{eqnarray}
where \bigskip $\Phi _{B}$ are the un-oscillated $^{8}B$ $\nu _{e}$ flux and 
$P_{\nu _{e}\rightarrow \nu _{i}}$ is the probability distribution between
the $\nu _{e}\rightarrow \nu _{i}$ transition. If $\rho _{NC}$ and $\rho
_{\mu ,\tau }$ satisfy the relation 
\[
\rho _{NC}|_{T_{NC}^{th}=5\text{MeV}}=\rho _{\mu ,\tau }|_{T_{\mu ,\tau
}^{th}=6.8\text{MeV}}\ , 
\]
then one can determine $\int dE\rho _{NC}P_{\nu _{e}\rightarrow \nu _{x}}$, $%
\int dE\rho P_{\nu _{e}\rightarrow \nu _{e}}$ and $\Phi _{B}$ provided $%
L_{1,A}$ is given. Unfortunately, the above relation, which implies $\rho
_{NC}|_{T_{NC}^{th}=5\text{MeV}}=\rho _{CC}|_{T_{CC}^{th}=5\text{MeV}}$,
does not hold, as shown in Fig. 1(a) in Appendix A.

\section{Conclusions}

We have analyzed the SNO and SK data on CC, NC and ES reactions to constrain
the leading axial two-body current $L_{1,A}.$ This two-body current
contributes the biggest uncertainty in every low energy weak interaction
deuteron breakup process, including SNO's CC and NC reactions. The only
assumption made in this analysis is that the total flux of active neutrinos
has the standard $^{8}B$ spectral shape (but distortions in the electron
neutrino spectrum are allowed). We have confirmed that SNO's conclusions
about the inconsistency of the no-flavor-conversion hypothesis and the
magnitude of the active solar neutrino flux do not have significant
theoretical model dependence. Our method has shown that SNO can be
self-calibrated or be calibrated by SK with respect to theoretical
uncertainties, and that the resulting calibration produces results in close
accord with theoretical expectations.  Alternatively, the purely experimental determination of \lia from reactor antineutrino data can be used to remove the dependence on theory, and SNO's conclusions are unaffected.



\vskip1 in \centerline{\bf ACKNOWLEDGMENTS} We would like to thank Petr
Vogel for useful discussions. JWC\ is supported, in part, by the Department
of Energy under grant DOE/ER/40762-213. RGHR is supported by the DOE under
Grant DE-FG03-97ER41020.

\appendix

\section{Computation of the Normalized Response Functions to the $^{8}B$
Spectrum}

In this Appendix, we define of the effective cross sections and then show
the numerical results that support eqs.(\ref{s}-\ref{sss}).

For CC and ES reaction, the effective cross section $\widetilde{\sigma }$ is
related to the true cross section $\sigma $ through the relation 
\begin{equation}
\widetilde{\sigma }_{CC(ES)}=\eta _{CC(ES)}\int_{T_{CC(ES)}^{th}}dT\int
dT^{\prime }r_{CC(ES)}(T,T^{\prime })\frac{d\sigma _{CC(ES)}}{dT^{\prime }}\
,
\end{equation}
where $\eta $ is an experimental efficiency, $T^{\prime }$ is the true
kinetic energy of the final state lepton and $T$ is the electron energy
recorded through the Cherenkov radiation of the electron, and $T^{th}$ is
the detection threshold. If the resolution of the detector were perfect, the
resolution function $r(T,T^{\prime })$ would be a delta function. For SNO, $%
r(T,T^{\prime })$ is a Gaussion function \cite{SNO1} 
\begin{equation}
r_{CC(ES)}(T,T^{\prime })=\frac{1}{\sqrt{2\pi }\Delta _{CC(ES)}}\exp \left[ -%
\frac{(T-T^{\prime })^{2}}{2\Delta _{CC(ES)}^{2}}\right]  \label{2}
\end{equation}
with resolution 
\begin{equation}
\Delta _{CC}=\left( -0.0684+0.331\sqrt{\frac{T^{\prime }}{\text{MeV}}}+0.0425%
\frac{T^{\prime }}{\text{MeV}}\right) \text{MeV\ .}
\end{equation}
For ES reactions, we have used the SK result rather than the SNO result for
better statistics. The resolution for SK is \cite{VFL} 
\begin{equation}
\Delta _{ES}=\sqrt{\frac{T^{\prime }}{10\text{MeV}}}1.5\text{MeV\ .}
\end{equation}

For NC reaction, the final state neutrons are thermalized then captured by
deuterons to form tritons and photons in SNO's first phase running. The
photons subsequently excite electrons which produce Cherenkov radiation.
Thus the SNO's NC events can be recorded as electron detections as well.
However, the kinematic information of the final state neutrons is lost in
thermalization. Thus the resolution function is monoenergetic \cite{NC} 
\begin{equation}
r_{NC}(T)=\frac{1}{\sqrt{2\pi }\Delta _{NC}}\exp \left[ -\frac{(T-T_{NC})^{2}%
}{2\Delta _{NC}^{2}}\right] \ ,  \label{3}
\end{equation}
where $T_{NC}=5.08$ MeV and $\Delta _{NC}=1.11$ MeV. In contrast to eq.(\ref
{2}), the effective NC differential cross section versus $E$ is not
distorted 
\begin{equation}
\widetilde{\sigma }_{NC}=\eta _{NC}\sigma
_{NC}\int_{T_{NC}^{th}}dTr_{NC}(T)\ .  \label{100}
\end{equation}

We now turn to the computation of the normalized response functions of the $%
^{8}B$ spectrum define in eq.(\ref{a4}). In Fig. 1(a), $\rho
_{CC}|_{T_{CC}^{th}=5\text{MeV}}$ (solid curve) and $\rho
_{NC}|_{T_{NC}^{th}=5\text{MeV}}$ (dashed curve) are shown as functions of $%
E $. The two curves are quite different. $\rho _{NC}$ is independent of $%
T_{NC}^{th}$, according to eqs.(\ref{a4}) and (\ref{100}). In contrast, the
peak of $\rho _{CC}$ can be shifted towards the high energy end by
increasing $T_{CC}^{th}$. When $T_{CC}^{th}=5$MeV, $\rho _{CC}$ is close to
a Gaussian function. Likewise, when $T_{e}^{th}=6.8$MeV, $\rho _{e}$ and $%
\rho _{\mu ,\tau }$ are adjusted to be close to Gaussians as well. Thus $%
\rho _{CC}$, $\rho _{e}$ and $\rho _{\mu ,\tau }$ can be related. To see how
different they are, it is convenient to define the following functions, 
\begin{eqnarray}
\delta \rho _{CC} &\equiv &\rho _{CC}|_{L_{1,A}=5\ \text{fm}^{3}}-\rho
_{CC}|_{L_{1,A}=0}\ ,  \nonumber \\
\delta \rho _{e} &\equiv &\rho _{e}|_{T_{e}^{th}=6.8\text{MeV}}-\rho _{\mu
,\tau }|_{T_{e}^{th}=6.8\text{MeV}}\ ,  \label{defs} \\
\delta \rho _{CC,ES} &\equiv &\rho _{CC}|_{T_{CC}^{th}=5MeV}-\rho
_{e}|_{T_{e}^{th}=6.8\text{MeV}}\ .  \nonumber
\end{eqnarray}
$\delta \rho _{CC}$, $\delta \rho _{e}$ and $\delta \rho _{CC,ES}$ are shown
as functions of $E$ in Fig. 1 (b)-(d), respectively.

\begin{figure}[!t]
\centerline{{\epsfxsize=5.0 in \epsfbox{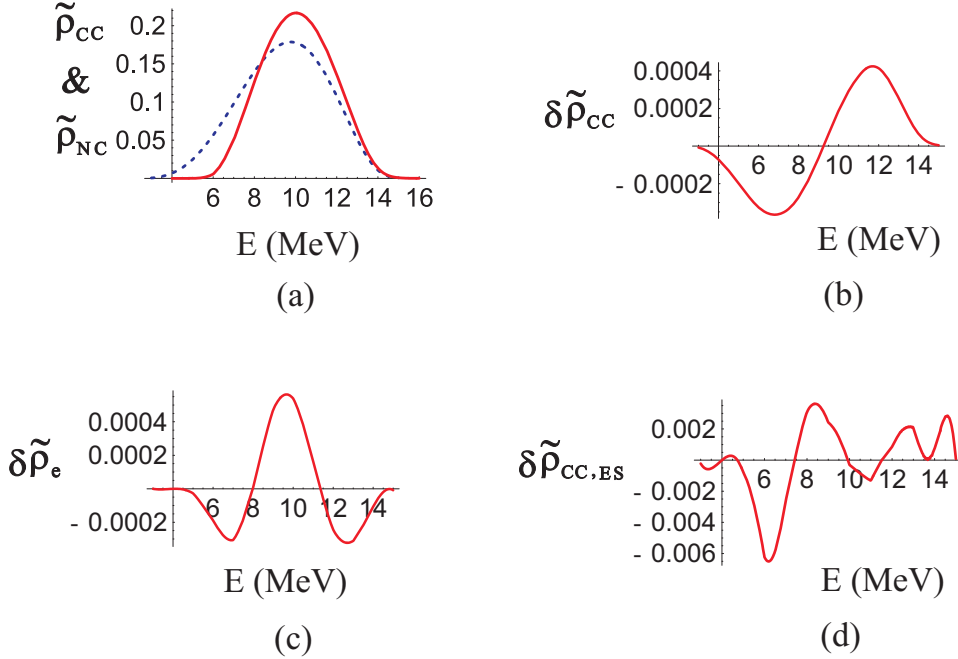}}}
\caption{{\it (a) $\protect\rho_{CC}|_{T_{CC}^{th}=5\text{MeV}}$ (solid
curve) and $\protect\rho_{NC}|_{T_{NC}^{th}=5\text{MeV}}$ (dashed curve) are
shown as functions of $E$. (b)-(d) $\protect\delta \protect\rho_{CC}$, $%
\protect\delta \protect\rho _{e}$ and $\protect\delta \protect\rho_{CC,ES}$
defined in eq.(\ref{defs}) are shown as functions of $E$, respectively. }}
\label{Fig:phaseshifts}
\end{figure}
%

To study the contributions of non-zero $\delta \rho $, we will first prove
an equality. Defining 
\begin{eqnarray}
\delta \rho _{\pm } &=&\frac{\delta \rho \pm \left| \delta \rho \right| }{2}%
\ ,  \nonumber \\
I_{\pm } &=&\int dE\delta \rho _{\pm }P_{\nu _{e}/\nu _{x}}\ ,
\end{eqnarray}
then 
\begin{equation}
\int dE\delta \rho P_{\nu _{e}/\nu _{x}}=I_{+}+I_{-}\ .
\end{equation}
Since 0$\leq P_{\nu _{e}/\nu _{x}}\left( E\right) \leq 1$, 
\begin{eqnarray}
0 &\leq &I_{+}\leq \int dE\delta \rho _{+}\ , \\
\int dE\delta \rho _{-} &\leq &I_{-}\leq \ 0\ .  \nonumber
\end{eqnarray}
Because $\int dE\delta \rho =0$, $\int dE\delta \rho _{\pm }=\pm \frac{1}{2}%
\int dE\left| \delta \rho \right| $. Thus we find 
\begin{equation}
\left| \int dE\delta \rho P_{\nu _{e}/\nu _{x}}\right| \leq \frac{1}{2}\int
dE\left| \delta \rho \right| \ .
\end{equation}
This model independent relation gives 
\begin{eqnarray}
\left| \int dE\delta \rho _{CC}P_{\nu _{e}/\nu _{x}}\right| &\leq &0.0013\ ,
\nonumber \\
\left| \int dE\delta \rho _{ES}P_{\nu _{e}/\nu _{x}}\right| &\leq &0.0012\ ,
\\
\left| \int dE\delta \rho _{CC,ES}P_{\nu _{e}/\nu _{x}}\right| &\leq
&0.0105\ ,  \nonumber
\end{eqnarray}
in comparison with $\int dE\rho P_{\nu _{e}/\nu _{x}}\sim 0.25$. The first
two inequality show that the $L_{1,A}$ dependence in $\rho _{CC}$ and the
difference between $\rho _{e}$ and $\rho _{\mu ,\tau }$ are negligible
compared with the correction from $\delta \rho _{CC,ES}$. The $\delta \rho
_{CC,ES}$ effect is parametrized by $\epsilon $ in eq.(\ref{m3c}) with 
\begin{equation}
\left| \epsilon \right| <4\%\ .
\end{equation}

\end{document}